\documentclass[twocolumn,showpacs,superscriptaddress]{revtex4-1}
\usepackage{hyperref,color}
\usepackage{graphicx}
\usepackage{bm}

\usepackage{amsthm,amsfonts,amssymb,epsfig,graphics,amsmath,oldgerm,color}

\usepackage[latin1]{inputenc}

\newcommand{\R}{{\mathbb R}}

\newcommand{\bspm}{\left(\begin{smallmatrix}}\newcommand{\espm}{\end{smallmatrix}\right)}
\newcommand{\bpm}{\left(\begin{matrix}}\newcommand{\epm}{\end{matrix}\right)}

\newcommand{\PROOF}{\textbf{Proof.} }

\def\epsilon{\varepsilon}
    
\def\beq{\begin{equation}}
\def\eeq{\end{equation}}

\newcommand{\subfigimg}[3][,]{%
  \setbox1=\hbox{\includegraphics[#1]{#3}}
  \leavevmode\rlap{\usebox1}
  \rlap{\hspace*{5pt}\raisebox{\dimexpr\ht1-.1\baselineskip}{#2}}
  \phantom{\usebox1}
}

\begin{document}


\title{Nonlinear Excitations in Magnetic Lattices with Long-Range Interactions}

\author{Miguel Moler\'on}
\affiliation{Institute of Geophysics, Department of Earth Sciences, ETH Zurich, 8092 Zurich, Switzerland}
\author{C. Chong$\,^*$}
\affiliation{Department of Mathematics, Bowdoin College, Brunswick, Maine 04011, USA}
\author{Alejandro J. Mart\'inez}
\affiliation{Oxford Centre for Industrial and Applied Mathematics, Mathematical Institute, University of Oxford,
Oxford OX2 6GG, UK}
\author{Mason A. Porter}
\affiliation{Department of Mathematics, University of California, Los Angeles, CA 90095, USA}
\author{P. G. Kevrekidis}
\affiliation{Department of Mathematics and Statistics, University of Massachusetts, Amherst, MA, 01003, USA}
\author{Chiara Daraio}
\affiliation{Division of Engineering and Applied Science California Institute of Technology
Pasadena, CA 91125, USA}

\date{\today}

\begin{abstract}

We study
 --- experimentally, theoretically, and numerically --- 
nonlinear excitations in lattices of magnets with long-range interactions. We examine breather solutions, which are spatially localized and periodic in time, in a chain with algebraically-decaying interactions. 
It was established two decades ago [S.~Flach, \emph{Phys. Rev. E} \textbf{58}, R4116 (1998)] that lattices with long-range interactions can have breather solutions in which the spatial decay of the tails has a crossover from exponential to algebraic decay. In this Letter, we revisit this problem in the setting of a chain of repelling magnets with a mass defect and verify, both numerically and experimentally, the existence of breathers with such a crossover. 


\end{abstract}



\keywords{Nonlinear Lattices, Fermi--Pasta--Ulam--Tsingou model, Breather, Long-Range Interactions, Magnets}
\maketitle

\emph{Introduction.} There has been considerable progress in understanding localization in nonlinear lattices over the past three decades~\cite{pgk:2011}. A prototypical example are spatially localized and temporally periodic discrete breathers (or just ``breathers'') \cite{Flach2007}. The span of systems in which breathers have been studied is broad and diverse; they include optical waveguide arrays and photorefractive crystals~\cite{moti}, micromechanical cantilever arrays~\cite{sievers}, Josephson-junction ladders~\cite{alex,alex2}, layered antiferromagnetic crystals~\cite{lars3,lars4}, halide-bridged transition metal 
complexes~\cite{swanson}, dynamical models of the DNA double strand \cite{Peybi}, Bose--Einstein condensates in optical lattices~\cite{Morsch}, and many others. Many of these studies concern models with coupling between elements only in the form of nearest-neighbor interactions.  However, there has been a great deal of theoretical and computational work in lattices with interactions beyond nearest neighbors. For example, some models of polymers \cite{Hennig2001}, quantum systems \cite{Choudhury96}, and optical waveguide arrays \cite{Kevrekidis2003,Christodoulides2002} have included
interactions beyond nearest neighbors; see also~\cite{PanosNNN,k13}. Dynamical lattices with long-range interactions (e.g., with all-to-all coupling) have been used as models for energy and charge transport in biological molecules \cite{Mingaleev1999}, and studies of such long-range models have explored phenomena such as equilibrium relaxation \cite{long1}, thermostatistics \cite{long2}, chaos \cite{long_chaos,long_chaos2}, and energy thresholds \cite{threshold,Flach_long}. Oscillators of numerous varieties have also been coupled via long-range interactions on lattices (and more general network structures) \cite{porter2016dynamical,arenas-review}. In fact, until recently, they were often assumed to be a fundamental ingredient for the formation of so-called ``chimera states''~\cite{knob1,knob2,abrams2015}.


Long-range interactions can have a significant effect on nonlinear excitations and yield phenomena that are rather different from those that result from only nearest-neighbor coupling. For example, stationary solitary waves with a nontrivial phase can arise both in discrete nonlinear Schr\"odinger (DNLS) equations with next-nearest-neighbor (NNN) interactions \cite{PanosNNN,ChongNNN} and in NNN discrete Klein--Gordon (KG) \cite{multibreathers_NNN} equations, and bistability of solitary waves is 
possible in DNLS equations with long-range interactions \cite{Rasmussen1997,Rasmussen1998}. Finally, most relevant for the present paper, breathers in KG and 
Fermi--Pasta--Ulam--Tsingou (FPUT) lattices with long-range interactions can exhibit a crossover from exponential decay (at short distances from the breather center) to algebraic decay (at long distances) if the interactions decay significantly slowly (specifically, algebraically slowly)~\cite{Flach_long}. A variety of new studies continue to elucidate fascinating consequences of long-range interactions.
For example, recent studies have revealed
the emergence of traveling discrete breathers without tails in nonlinear lattices with suitable
long-range interactions~\cite{doi1} and the emergence of a linear
spectral gap, which enables the emergence of
a low-frequency breather~\cite{doi2}, in nonlinear lattices with other long-range interactions.
Although there are many theoretical and computational studies of lattice systems with long-range interactions, we are not aware of any experimental realizations of breathers in such systems.



In this paper, we use experiments, theory, and numerical computations to study a strongly nonlinear lattice with long-range interactions that decay algebraically. Specifically,
we consider 
a one-dimensional (1D) chain of repelling magnets with a single mass defect.
This system allows us to realize fundamental structures, such as solitary waves, in a tabletop setup with
real-time spatio-temporal resolution ~\cite{moleron,Mehrem2017}. 
Moreover, the use of magnetic interactions allows exciting applications. They have already been used as a passive mechanism to couple nodes of a lattice for unidirectional wave-guiding \cite{Neel}; and it has been suggested that magnetic interactions can be used to
design novel devices for frequency conversion \cite{Marc2017} and shock absorption \cite{moleron}. 
In our study, we focus on breathers in a magnetic chain and demonstrate that there is a crossover from exponential decay to algebraic decay in the spatial profile of these breathers.
Our numerical findings are consistent with theoretical predictions
that were developed almost twenty years ago in~\cite{Flach_long}, 
and they agree quantitatively with our experiments.

 \begin{figure}[h!]
\centering
\includegraphics[width= .8 \linewidth]{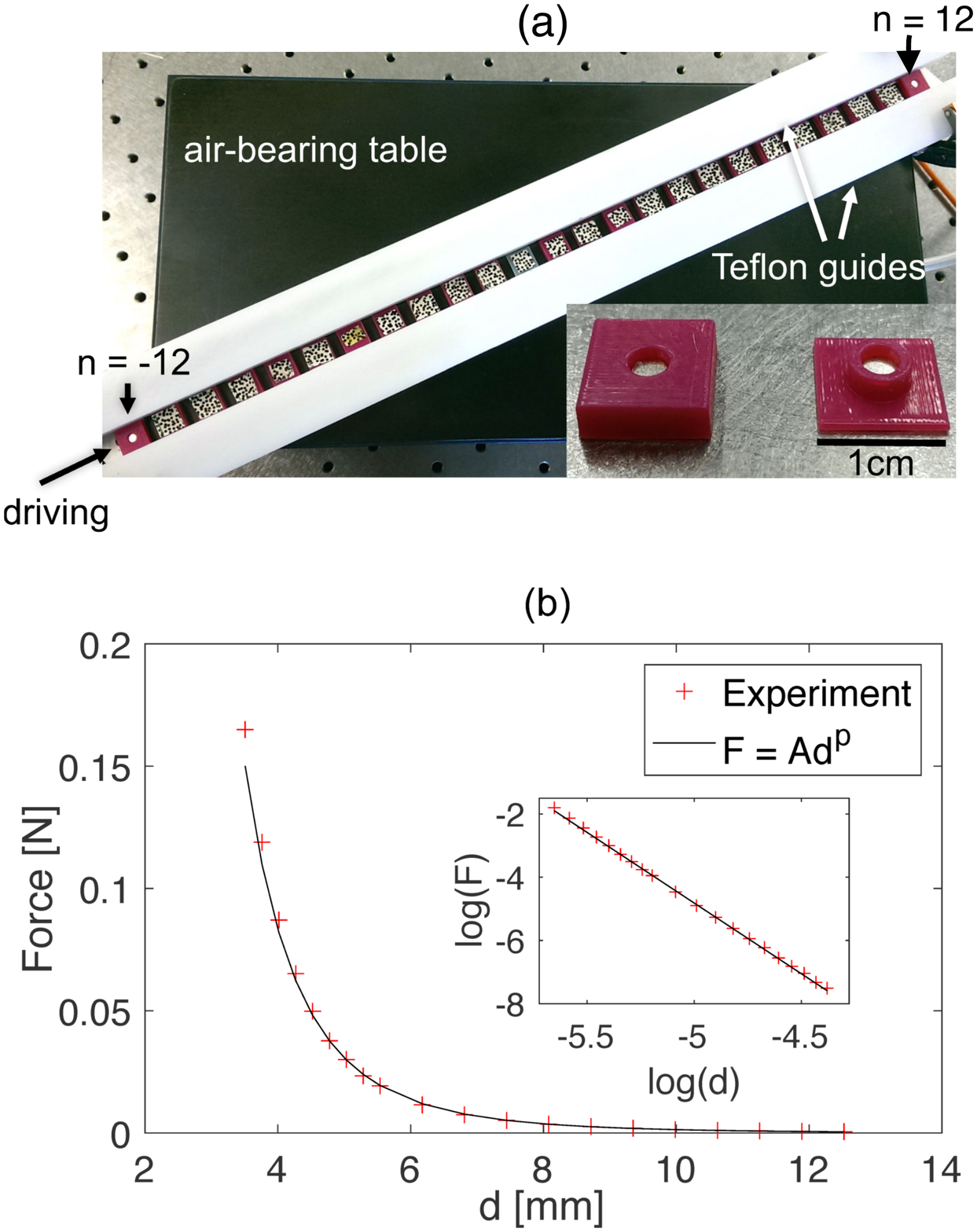}
\caption{(a) Picture of our experimental setup. The lattice consists of 25 magnetic particles deposited on an air-bearing table. The right boundary ($n=12$) is fixed, and the left boundary ($n=-12$) is 
driven harmonically with an electrodynamic transducer. The magnetic particles are composed of a disc magnet (type Supermagnete \href{https://www.supermagnete.de/eng/data_sheet_S-03-01-N.pdf}{S-03-01-N}, with magnetization grade N48, a diameter of 3~mm, and a height of 1~mm). 
The inset shows a magnified view of the magnetic particles embedded in a 3D-printed support: (left) normal particle and (right) defect particle. (b) Relationship between the force $F$ and center-to-center separation distance $d$ between two particles. The plus signs represent experimental data, and the solid curves represents 
a curve of the form $F=Ad^p$. 
In the inset, we show a plot of $\log(F)$ versus $\log(d)$ that we use for fitting the exponent $p$ and the magnetic coefficient $A$. 
}
\label{Fig_Exp_Setup}
\end{figure}

\emph{Experimental setup.} In Fig.~\ref{Fig_Exp_Setup}(a), we show a picture of our experimental setup. We situate an array of disc magnets over a 150~mm$\times$300~mm rectangular air-bearing table from \href{http://ibspe.com/product/air-bearings-flat-rectangular-150mm-x-300mm.htm}{IBS Precision Engineering} (to reduce surface friction) and 
between two Teflon rectangular rods (to restrict the particle motion to one dimension). As shown in the inset of Fig.~\ref{Fig_Exp_Setup}(a), we insert each magnet into a 3D-printed support. We glue a glass slide below the 3D-printed support to obtain a desired amount of levitation. The magnets are axially magnetized, and they have the same orientation, so each magnet repels its neighbors. The average mass of the non-defect  
particles in the 25-particle chain is $M=0.45$~g (with a standard deviation of $s=0.0028$), and the mass of the defect particle is $m=0.20$~g. 
To excite the chain harmonically, we glue the left boundary to an aluminum bar attached to an electrodynamic transducer (\href{https://www.beyma.com/getpdf.php?pid=5MP60/N}{Beyma 5MP60/N}). 
The measured total harmonic distortion of this transducer is 
below $10\%$ in the amplitude range (between 0 and 4 cm) under consideration.

%
We measure the motion using a digital image correlation (DIC) software from Correlated Solutions (\href{http://correlatedsolutions.com/vic-2d/}{VIC 2D}). 
We use a camera (of model \href{https://www.ptgrey.com/grasshopper3-41-mp-color-usb3-vision-cmosis-cmv4000-2-camera}{GS3-U3-41C6C-C} from Point Gray) to record the particles' motion at a frame rate of 200 fps. 
To help track the particles, we glue speckle patterns to the top of the 3D-printed support [see Fig.~\ref{Fig_Exp_Setup}(a)]. We postprocess the video files with the VIC software to extract particle displacements and velocities. As in \cite{moleron,Neel}, we assume that the relationship between the repelling force and distance has the form $F=Ad^p$, where $F$ is the force and $d$ is the center-to-center separation distance between two particles. 
We estimate the magnetic coefficient $A$ and exponent $p$ by measuring the repelling force at 22 separation distances [represented by plus signs in Fig.~\ref{Fig_Exp_Setup}(b)]. We measure the repelling force by fixing one magnet to a load cell 
(of type \href{https://www.omega.com/pressure/pdf/LCL.pdf}{OMEGA LCL-113G}) and approaching another magnet using a high-precision translation stage. Using a least squares fitting routine for $\log(F)$ versus $\log(d)$ with our experimental data [see the inset in 
Fig.~\ref{Fig_Exp_Setup}(b)] yields $A \approx 1.5683\times~10^{-12}N/m^p$ and $p \approx -4.473$. We use these parameter values throughout the text.
%



\emph{Theoretical Setup.} Our experimental setup motivates the following model (which assumes that each node, representing a magnet, is coupled to every node in a chain): 
\begin{align} \label{model} 
	 M_n \ddot{u}_n =  & \sum_{j = 1}  A \left( j \delta_0 + u_n - u_{n-j} \right)^{p} \\ \nonumber 
	 	& -  A \left( j \delta_0 + u_{n+j} - u_n \right)^{p} - \eta \dot{u}_n\,,
\end{align}
where 
$u_n=u_n(t)\in\R$ is the displacement of the $n$th magnet from its equilibrium position, 
 the mass of the $n$th magnet is $M_n$, the magnetic coefficient is $A$, and the nonlinearity exponent is $p$
 (Fig.~\ref{Fig_Exp_Setup}(b) shows the spatial decay in the force with respect to the center-to-center distance between particles).
 This model assumes that each magnet, including its magnetic properties, 
  is identical.
 The equilibrium separation distance between two adjacent magnets in an infinite lattice is $\delta_0$. In a finite lattice,
 the equilibrium separation distance will depend on the lattice location, see the Supplementary Material for details.
 We model damping effects with a dashpot term $\eta \dot{u}_n$, where we empirically estimate the damping factor $\eta$ (see our discussion below). We apply a harmonic boundary drive $u_{\rm left }(t) = a \sin( 2 \pi f_b t)$, where $a$ denotes the drive amplitude and $f_b$ denotes its frequency. Our initial theoretical considerations involve a Hamiltonian lattice, so we take $a=\eta=0$. Later, when we compare our numerical results to experiments, we also consider nonzero values of the drive amplitude and damping factor.
%
%

\begin{figure}[h!]
\centering
\includegraphics[width= .75 \linewidth]{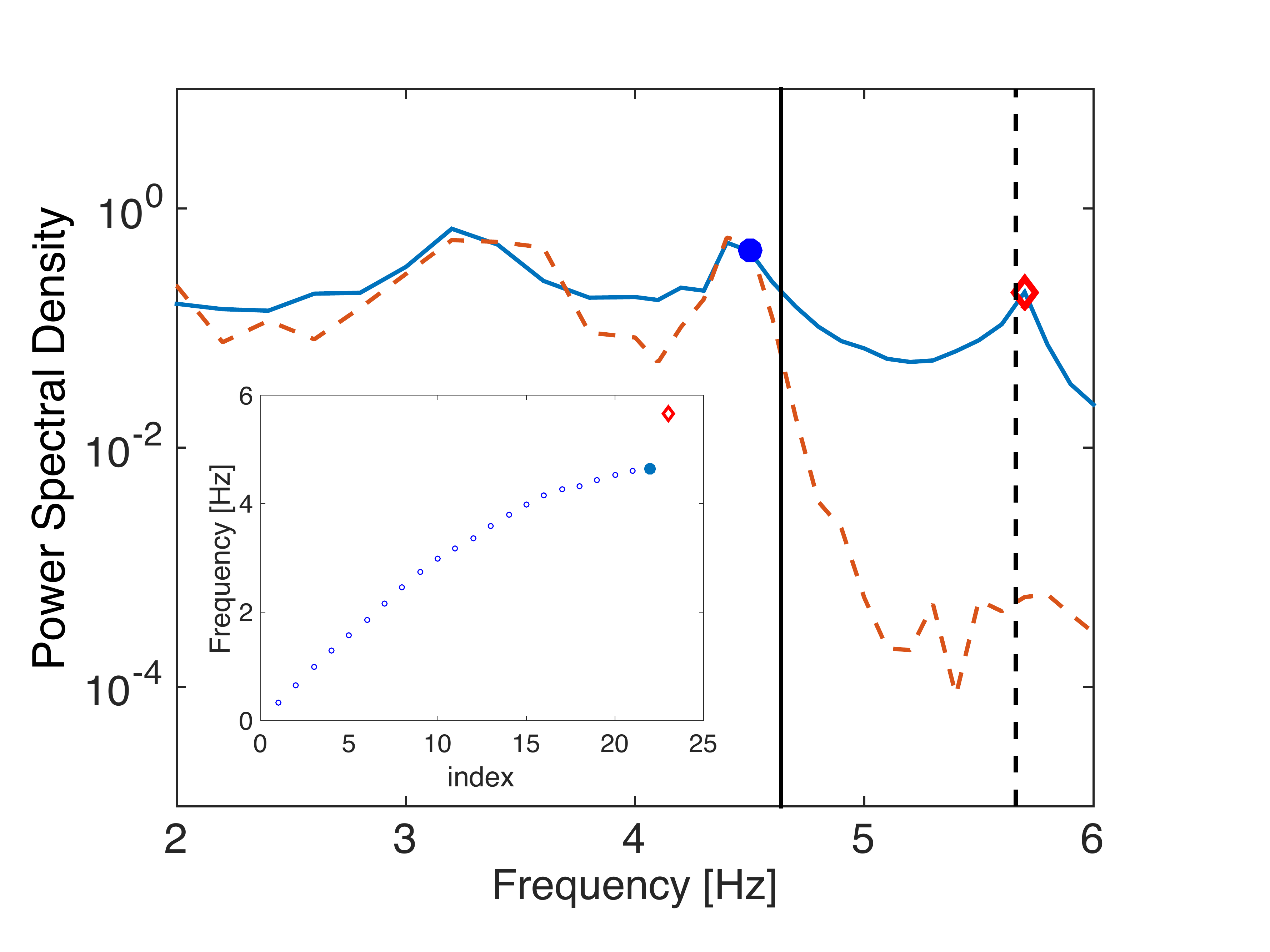}
\caption{Experimental power spectral density (PSD) for a homogeneous chain of 25 magnets (dashed red curve) and
for a magnetic chain with a mass defect located at site $n_d = -8$ (solid blue curve). The experimental cutoff frequency (blue disk) and defect mode (red diamond) are in reasonable agreement with the numerical prediction (vertical solid and dashed lines). 
In the inset, we show numerically computed eigenfrequencies for a chain with 23 particles (25 including the two fixed boundaries)
and a single
defect at site $n_d = -8$ and no damping or driving ($\eta = a =0$). 
The blue disk in the inset represents the numerical cutoff frequency,
and the red diamond shows the numerical defect mode.
} 
\label{fig:disp}
\end{figure}

In a homogeneous chain (where all masses are identical, so $M_n = M$) the linearization of \eqref{model} has plane-wave solutions $u_n = \exp( i k n + i\omega t)$, where
\begin{align} \label{eq:disp}
	\omega^2(k) &= K_2  \sum_{j=1}^\infty \frac{1}{j^s}[1 - \cos( j k ) ] \\ 
		&= K_2 [   \zeta(s) - \text{Re}\{e^{i k} \phi(e^{i k},s,1) \} ]\,, \nonumber
\end{align}
where $s=1-p$, the linear stiffness is $K_2 = - 2 A p \delta_0^{p-1}/M$, the Riemann zeta function is $\zeta(s)$, and $\phi(z,s,a)$ 
is the Hurwitz--Lerch transcendent function \cite{DLMF}.
This dispersion curve is nonanalytic
in the wavenumber $k$, because its $\kappa$th derivative (where $\kappa$ is the integer satisfying $s-1 \leq \kappa < s$) with respect to $k$ is discontinuous at $k=0$. Below we discuss the consequences of this nonanalyticity. 
The dispersion curve is analytic at the upper band
edge (i.e., at $k=\pi$). 

Because we are interested in solutions that decay spatially to $0$ at infinity, it is natural to seek breather
frequencies that lie above the edge of the spectrum $\omega(\pi)$ (to avoid resonances with linear modes). Equation~\eqref{model} with $M_n = M$ is not an appropriate model for seeking small-amplitude (bright)
breather solutions, because one needs the plane waves to have a modulational instability, which is not
possible in a homogeneous magnetic chain \cite{Flach2007}.
Hence, to obtain breathers, we break the uniformity of the chain
 by introducing a light-mass defect, motivated by the analysis 
 of~\cite{Theocharis2009} for nonlinear lattices with nearest-neighbor interactions.
 This creates a defect mode that lies above the edge of the linear spectrum,
 from which breathers can bifurcate. Breathers in nearest-neighbor FPUT-like lattices with defects have been studied extensively
both theoretically \cite{Theocharis2009} and experimentally \cite{nature11}. To find breathers in a magnetic chain, one can alternatively use a lattice with spatial heterogeneity (e.g., a dimer)~\cite{Theocharis10,Boechler2010,Huang2}
or one with an on-site potential \cite{James2011,James2013} or local resonators \cite{Liu2015,Liu2016}.

A chain with a single mass defect is the starting point for our model with long-range interactions. We reduce the mass of the $n_d$th node (but without modifying its magnetic properties) by adjusting the support in which the magnet is embedded [see Fig.~\ref{Fig_Exp_Setup}(b)]. 
Consequently,
\begin{equation}
	M_n = \left \{
\begin{array}{cc}
	m\,, & 
	n  = n_d \\
	M\,, & \mbox{otherwise} \
\end{array} \right. \,,
\end{equation}
where $n_d$ is the index of the mass defect with mass $m< M \in \R$ and $M$ is the mass of the
non-defect nodes. In Fig.~\ref{fig:disp}, we show the spectrum of the linearized chain with a single mass defect.


 \begin{figure}
 \centerline{
  \epsfig{file=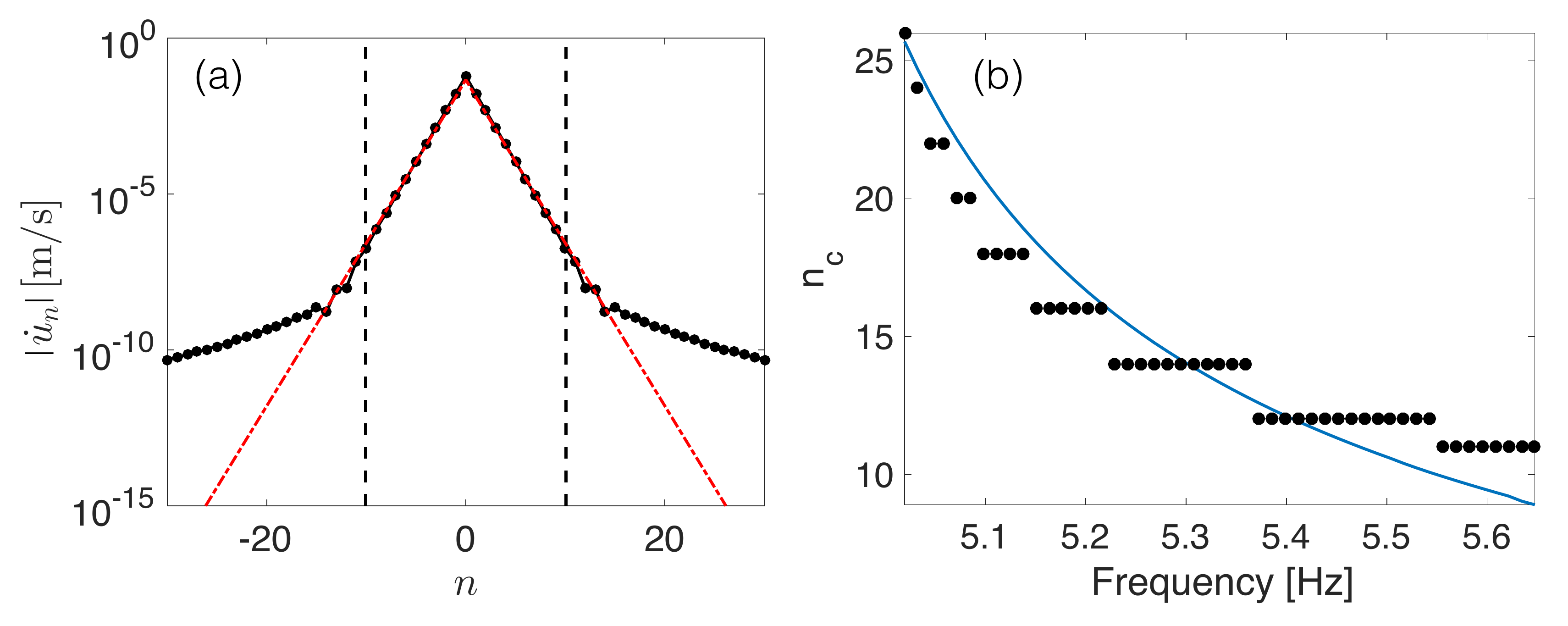,width= \linewidth} }
 \caption{(a) Semi-log plot of a breather solution (black curve with markers), with a frequency of $f_b \approx 5.54$ Hz, of Eq.~\eqref{model} with $\eta = a = 0$ for a magnetic chain with a defect particle in the center ($n_d = 0$). The vertical axis gives the absolute value of the velocity, and the horizontal axis gives the node index. 
  For comparison, we show a breather solution of the same frequency for a lattice with only nearest-neighbor interactions (red dash-dotted curve). The vertical dashed line is the predicted value of the crossover value $n_c$ 
 from Eq.~\eqref{eq:law}. (b) Numerically computed crossover point (black markers) and prediction based on Eq.~\eqref{eq:law} (curve).
 }
 \label{fig:trans}
\end{figure}


\emph{Numerical Results.} 
 We start by numerically computing time-periodic solutions of the Hamiltonian variant of Eq.~\eqref{model} (i.e., with $a=\eta=0$) and  $N=65$ nodes. The values that we use for the magnetic
potential parameters are $A \approx 1.5683\times~10^{-12}N/m^p$ and $p \approx -4.473$. Each particle, except for the defect in the center, has a mass of $M=0.45$~g; the mass of the defect node is $m=0.20$~g.  We numerically compute
the linear spectrum and obtain a defect mode with frequency $f_d \approx 5.66$Hz. We use this linear mode as an initial guess in a Newton method and identify a time-periodic solution with a frequency slightly below the defect frequency (see the Supplementary Material for details on numerical computations). 
In Fig.~\ref{fig:trans}(a), we show a semi-log plot of the absolute value of the velocity profile of the breather that we obtain using Newton iterations. One of the defining features of a breather in lattices with nearest-neighbor interactions is exponential decay of the tails. [See the red dashed curve in Fig.~\ref{fig:trans}(a).] The linear slope of the breather in the semi-log plot suggests that there is exponential decay of the tail close to the center. In fundamental contrast to its nearest-neighbor counterpart, the breather in the lattice with long-range interactions exhibits a transition at a critical lattice site $n_c$, and the decay becomes algebraic rather than exponential.
 This feature was first observed about two decades ago in a KG lattice 
with a cubic potential (i.e., in the $\phi_4$ model) \cite{Flach_long}, which has long-range interactions with coefficients with algebraic decay (in particular, they have a power-law decay $\mathcal{O}(1/n^{s})$ with respect to node $n$).
The linearization of Eq.~\eqref{model} also has interaction coefficients with power-law decay $\mathcal{O}(1/n^{s})$. 

The algebraic decay of the breather far away from its center arises as follows. Its amplitude is small away from its center, so we can linearize the equations of motion. Additionally, because the breather is temporally periodic, we can express the time dependence of the solution as a Fourier series $u_n(t) = \sum_j \hat{u}_n(j) e^{i j \omega_b t}$,
where $\omega_b = 2 \pi f_b$ is the breather's angular frequency. One computes the Fourier coefficients using Green's functions \cite{Flach_long} to obtain
\begin{equation} \label{eq:fourier}
	\hat{u}_n(j) = \int_0^{2 \pi} \frac{ \cos(k j)}{ (j \omega_b)^2 - \omega^2(k)} \, dk\,,
\end{equation}
where $\omega^2(k)$ is given by the dispersion relation in Eq.~\eqref{eq:disp}. Now it is clear why it is important to highlight the nonanalytic nature of $\omega^2(k)$: the Fourier coefficients in Eq.~\eqref{eq:fourier} with discontinuities in the $\kappa$th derivative yield Fourier series that converge algebraically. This implies that $u_n  \sim 1/n^s$ for large $n$ \cite{Flach_long}. One can make similar arguments to explain the exponential decay near the center. (See \cite{Flach_long} for details.)



Assuming that the proportionality constants of the exponential decay
and the algebraic decay are roughly the same, there is a crossover point between the two types of decay that satisfies $e^{\nu n_c} = \frac{1}{n_c^s}$, where $\nu$ is the exponential decay rate of the breather near the center. 
This yields the following prediction for the crossover site $n_c$ \cite{Flach_long}:
\begin{equation} \label{eq:law}
	 \frac{\log n_c}{n_c} = \frac{\nu}{1-p}\,.
 \end{equation}
For the solution in Fig.~\ref{fig:trans}(a), the predicted crossover is $n_c =10$, which is roughly where the decay properties change in the numerical solution [see Fig.~\ref{fig:trans}(a)]. Because we made several assumptions to derive
Eq.~\eqref{eq:law}, we also compute the crossover point from the numerically-obtained breather solutions. We calculate this point numerically by determining the first node at which the deviation of the solution from the best-fit line in the semi-log scale exceeds 1\%
of the solution amplitude. In the example in Fig.~\ref{fig:trans}(a), this yields a crossover point of $n_c =12$.
Equation~\eqref{eq:law} predicts that the crossover location depends on the solution's exponential decay rate $\nu$, which in turn depends on the breather frequency $f_b$. In Fig.~\ref{fig:trans}(b), we show a comparison of observed numerical crossovers 
and Eq.~\eqref{eq:law} for various breather frequencies.


 \begin{figure*}
 \centerline{
     \epsfig{file=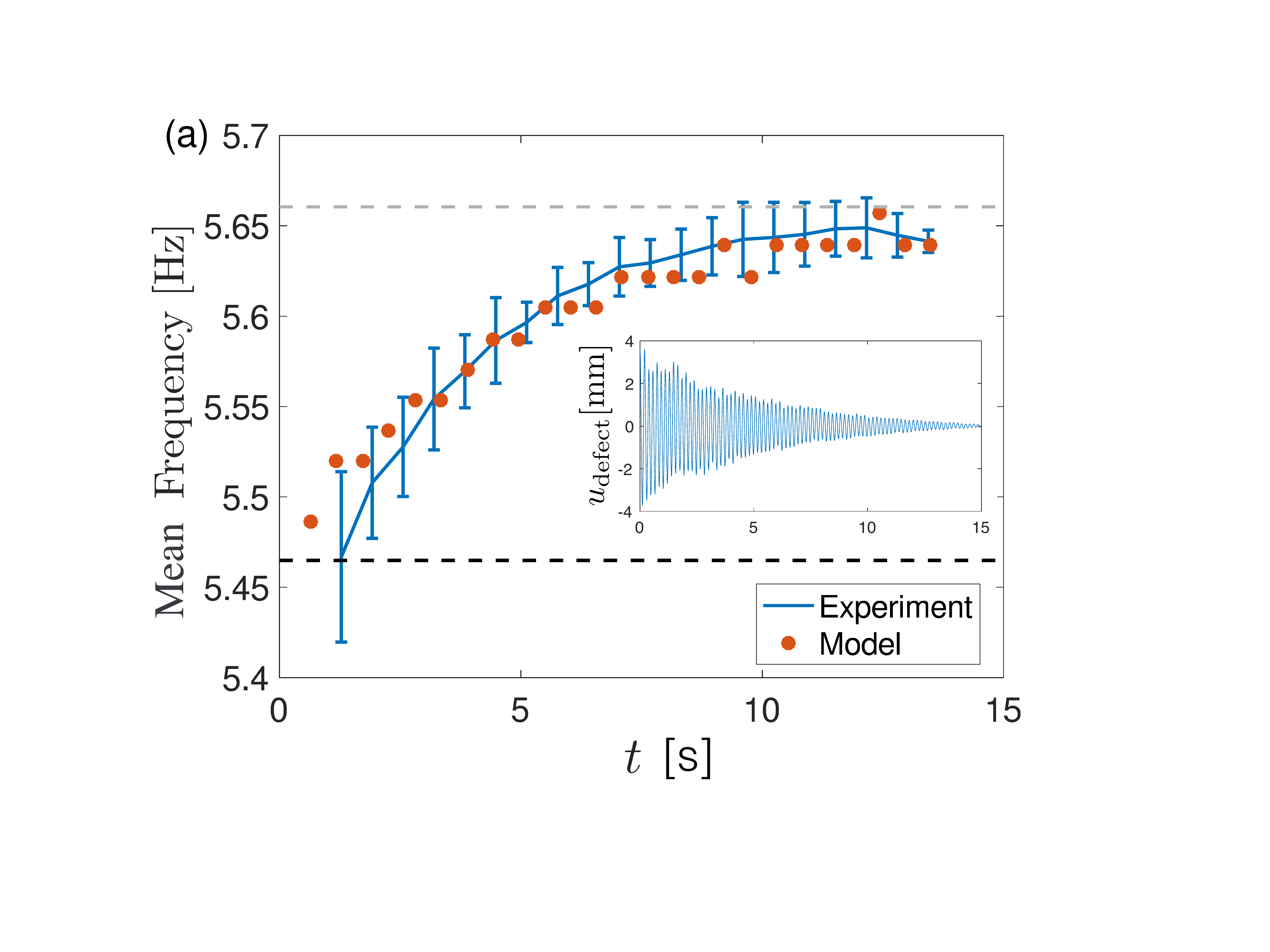,width= .33 \linewidth}
  \epsfig{file=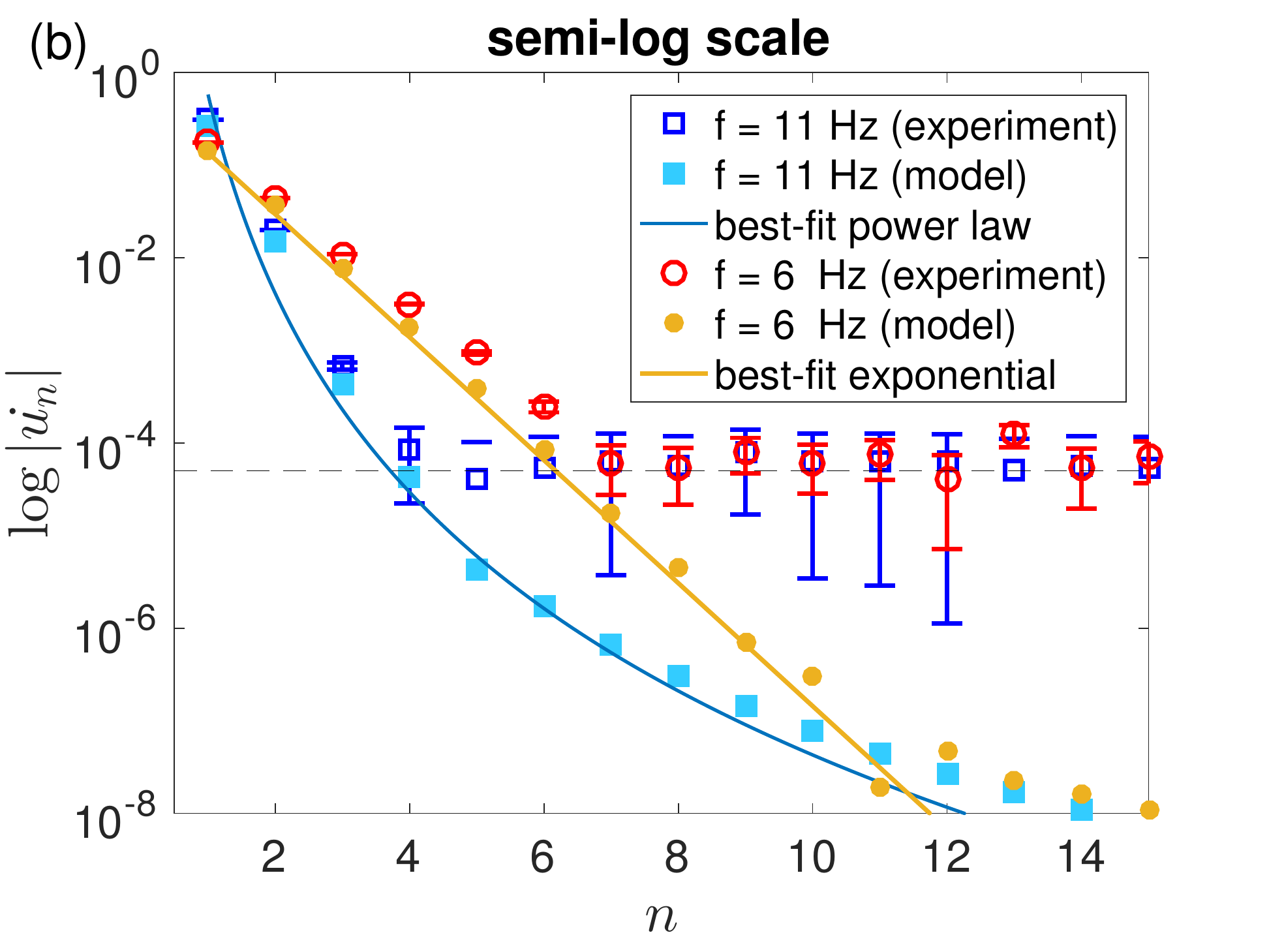,width= .33 \linewidth}
  \epsfig{file=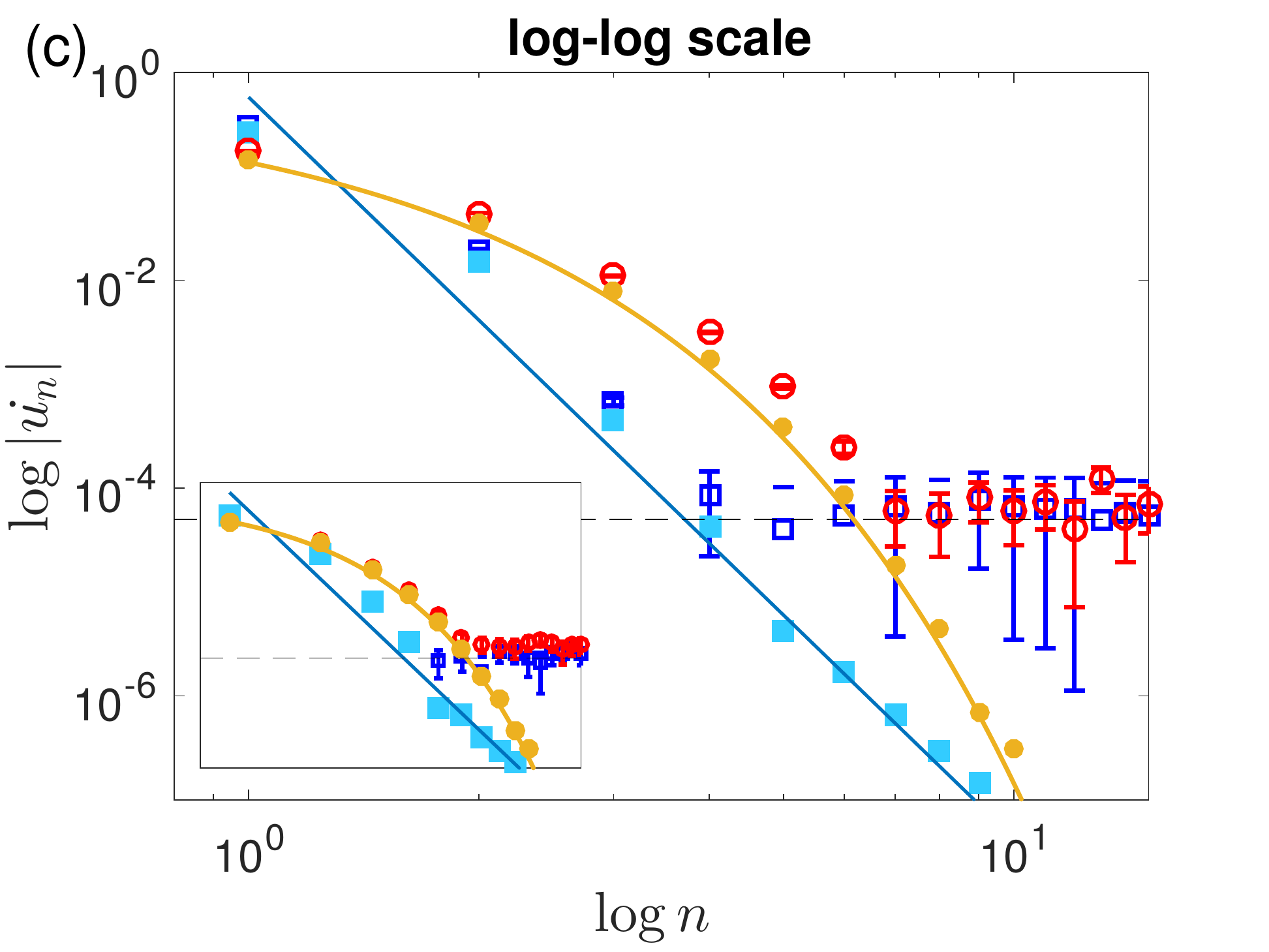,width= .33 \linewidth}
      }
 \caption{\textbf{(a)} Experiment initialized with a Hamiltonian breather solution
 of Eq.~\eqref{model} with frequency $f_b \approx 5.46$ Hz. We show the mean oscillation frequency of the defect particle for every $1.28$ s for the experiment (blue markers with error bars) and model with damping (with $\eta = 0.10$~g/s) but no driving (red disks). The error bars indicate the standard deviation over 5 experimental realizations. Note that the node 
 oscillates initially at the predicted frequency.
  The frequency approaches the sole defect frequency of the linear system, as the damping cause displacements to approach $0$. In the inset, we show an example of defect-particle decay from an experiment.
  \textbf{(b)} Semi-log plot of the experimental data for drive frequencies of $f_b = 6$ Hz (open red circles with error bars) and  $f_b = 11$ Hz (open blue squares with error bars).
  The chain is homogeneous (there is no defect particle), because the boundary drive is acting like the defect particle (which we label as $n=0$).
  We show our predictions from the damped, driven model (filled markers) as well as the best fit
 to exponential (yellow curve) and power-law (blue curve) decay. The experimental data for $f_b = 6$ Hz follows a roughly linear trend in the semi-log plot, suggesting that its decay is exponential.
  \textbf{(c)} Same as panel (b) but as a log-log plot. The experimental data for $f_b = 11$ Hz follows a roughly linear trend in the log-log plot, suggesting that its decay is algebraic. 
 Panels (b) and (c) share the legend that we show in (b). The inset in panel (c) shows a similar result for a chain of length $N=29$ (which has a smaller equilibrium distance). In this case, more nodes have 
 an amplitude that is comparable to the amount of noise.
 }
 \label{fig:compare}
\end{figure*}


\emph{Experimental Results.} 
We now turn our attention to the experimental realization of breathers in a nonlinear lattice with long-range interactions. For our experiments, we consider a chain of $N=25$  magnets (including the boundaries) with a defect magnet at site $n_d = -8$. We experimentally probe the linear spectrum
by performing a frequency sweep. To do this, we excite 
the chain at 33 frequencies between $2$ and $6$~Hz and extract the resulting steady-state displacement amplitudes at the excitation frequency in different locations. The red dashed line in Fig.~\ref{fig:disp} 
represents the power spectral density (PSD) of particles $-4$--$0$, and the blue solid line represents the PSD of the defect particle. 
The model prediction based on the Hamiltonian limit (with $\eta = a = 0$) of Eq.~\eqref{model} (which was computed numerically, as shown in the inset of Fig.~\ref{fig:disp}) agrees with 
the experimentally-observed passband cutoff frequency $f \approx 4.50$~Hz and defect-mode frequency $f_d \approx 5.66$~Hz.



To further evaluate our model, we initialize the experimental chain using the displacements that correspond to the theoretically-predicted Hamiltonian breather with frequency $f_b \approx 5.46$~Hz. The nodes oscillate initially with the predicted frequency [see Fig.~\ref{fig:compare}(a)].  In this particular experiment, we do not add energy to the system. 
Thus, as the oscillation amplitude decreases due to damping, the dynamics gradually becomes more linear and the oscillation frequency approaches the sole linear defect-mode frequency $f_d \approx 5.66$~Hz. 
We use this experiment to empirically determine the damping parameter $\eta = 0.10$~g/s to match the temporal amplitude decay of the defect particle. [See the inset in Fig.~\ref{fig:compare}(a).] 
We conduct an analogous numerical experiment using Eq.~\eqref{model} with damping but no driving (specifically, $\eta = 0.10$ g/s and $a=0$), which matches the observed experimental data; see the solid red disks in Fig.~\ref{fig:compare}(a).

Our final experiment probes the decay properties of the breather. To allow the experimental system to reach a steady state (which allows us to more closely examine these
properties), we again harmonically excite the left boundary magnet, so the displacement of the boundary magnet is $u_{\rm left} = a \sin( 2 \pi f_b )$. We thereby treat the boundary as a ``core" of the breather, so we do not use a defect particle in these experiments. We seek time-periodic solutions of Eq.~\eqref{model}
that account for both the boundary excitation and damping effects. We use the parameter values $\eta = 0.10$ g/s and $a = 3.8$ mm.  
The transition that we observe in Fig.~\ref{fig:trans}(a) occurs at
amplitudes, which we estimate to be $0.05$~mm/s, below the amount of noise in the experiments.
This value corresponds to the mean velocity amplitudes of particles $9$--$24$, whose 
motion can be attributed primarily to ambient vibrations. Thus, for the drive (breather) frequency $f_b = 6$ Hz, we observe only exponential decay. 

However, for a drive frequency of $f_b = 11$ Hz, the transition to algebraic decay
occurs close to the core of the breather, so there appears to be a glimpse of
the associated decay prior to reaching the level at which ambient noise vibrations overwhelm the algebraic tail. Note that the crossover approaches the core of the breather as the breather frequency increases [see Fig.~\ref{fig:trans}(b)]. In Figs.~\ref{fig:compare}(b,c), we show the tails of the breather in semi-log and log-log plots. 
For $f_b = 6$ Hz, the experimental data (open red circles with error bars) has a roughly 
linear trend in the semi-log plot, suggesting that its decay is exponential. The experimental data follows the model prediction (solid yellow circles) up to the point at which it reaches the noise level (the horizontal gray dashed line). We fit (using a least squares procedure) the model solution with an exponential curve of the form $\alpha e^{-\beta n}$ (solid yellow curve), and we obtain
$\alpha \approx 0.6287$ and $\beta \approx 1.529$. For $f_b = 11$ Hz,
the experimental data (open blue squares with error bars) has a roughly linear trend in a log-log plot, suggesting its algebraic decay.
The experimental data follows our model's prediction (closed light blue squares) until  reaching the noise level (horizontal gray dashed line). We fit the model solution with a power-law curve of the form $\alpha n^{-\beta}$ (solid blue curve),
and we obtain $\alpha \approx 0.579$ and $\beta \approx 7.131$.
Our results for other parameter values are similar. 
For example, in the inset of Fig.~\ref{fig:compare}(c), we show a log-log plot of periodic solutions with $f_b = 9$ Hz (red) and $f_b = 13$ Hz (blue) for a chain with $N=29$ nodes. Because the lattice 
is confined to a length of $L \approx 33.7$ cm, the equilibrium distance is about $6/7$ of the one
in the $N=25$ chain. This increases the linear stiffness and hence increases the passband cutoff. 
Consequently, we need higher frequencies to avoid resonance with the linear modes.


\emph{Discussion and Conclusions.} We studied 
a lattice of magnets with long-range interactions, and we obtained quantitative agreement between theory, numerics, and experiment. Specifically, using a combination of experiments, computation, and analysis, we explored
the prediction of \cite{Flach_long}, made about twenty years ago, that
the tail of a breather solution of this nonlinear lattice exhibits a transition from exponential to algebraic decay. As far as we are aware, our work represents the first experimental realization of breathers in a nonlinear lattice with long-range interactions.


The study of long-range interaction systems is an increasingly important topic in numerous and wide-ranging areas of physics. These include dipolar Bose--Einstein condensates (BECs)~\cite{lahaye}, where the
recent formation of quantum droplets and their bound states~\cite{pfau}
suggests that interesting types of long-range interactions can also arise in the study of BECs in optical lattices. Long-range interactions also play important roles in the study of coupled phase oscillators in diverse physical settings \cite{arenas-review}, heat transport in oscillator chains coupled to thermal reservoirs \cite{Olivares16,iubini2017}, and more.
Moreover, our experimental setup of magnet lattices has the potential to enable systematic, well-controlled studies of phenomena
involving these mixed (exponential and algebraic) decaying breathers.
For instance, it would be especially interesting to examine what happens when such breathers interact and how the decay properties (and interactions between breathers) depending on lattice dimensionality.


\emph{Acknowledgements.} This material is based upon work supported by the National Science Foundation under Grant No. 
DMS-1615037 (awarded to CC) and EFRI Grant No. 1741565 (awarded to CD).
AJM acknowledges support from CONICYT (BCH72130485/2013). PGK gratefully acknowledges support from the US-AFOSR via FA9550-17-1-0114. 

\bibliography{granular6}

\newpage

\onecolumngrid

{ \Huge
\centerline{ \bf Supplementary Material}
}

\section{Equations of Motion for Finite Chain}

\begin{figure}[h]
\centering
   \begin{tabular}{@{}p{0.4\linewidth}@{\quad}p{0.4\linewidth}@{\quad}@{}}
        \subfigimg[width=\linewidth]{\bf (a)}{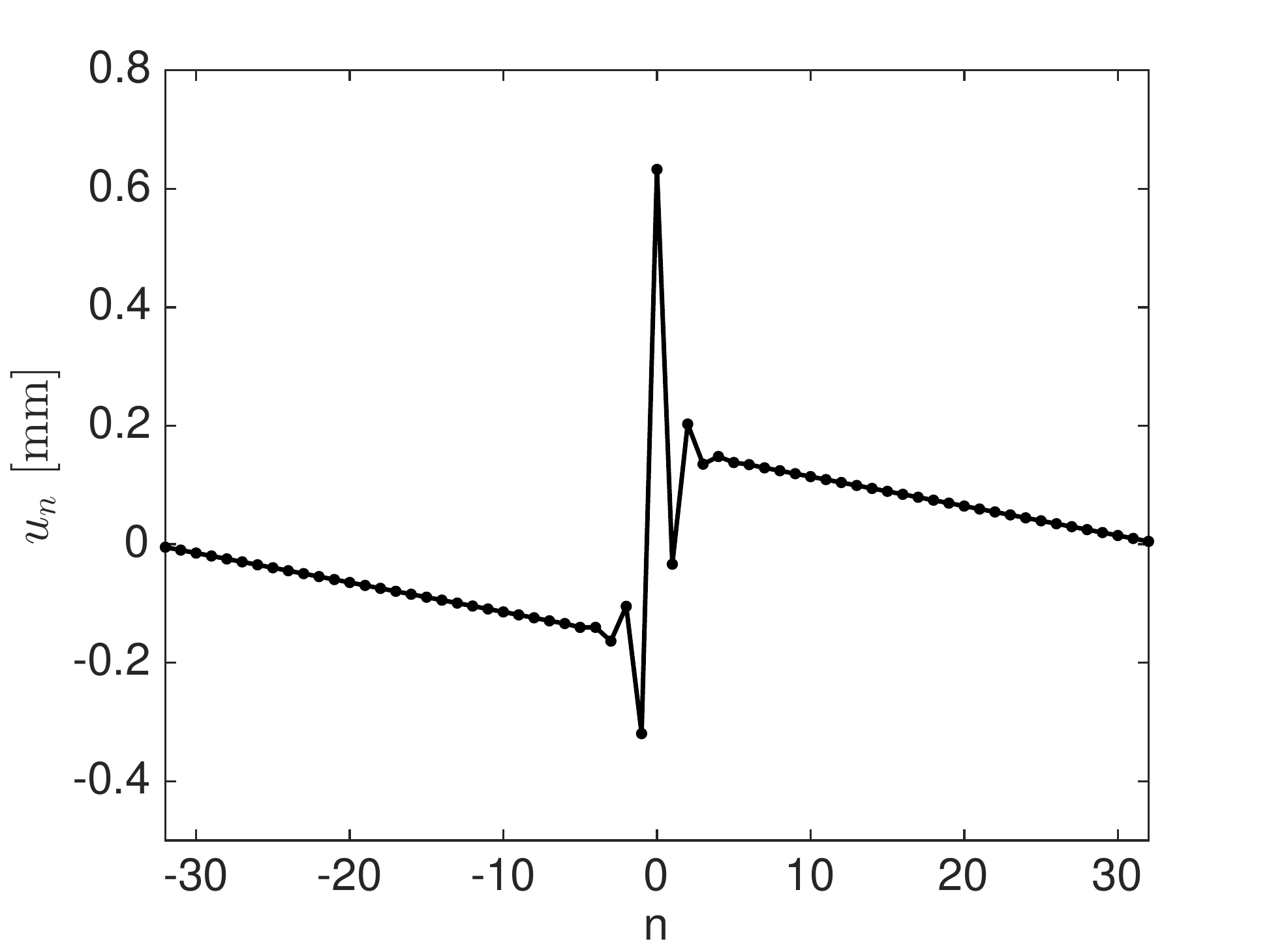}  &
    \subfigimg[width=\linewidth]{\bf (b)}{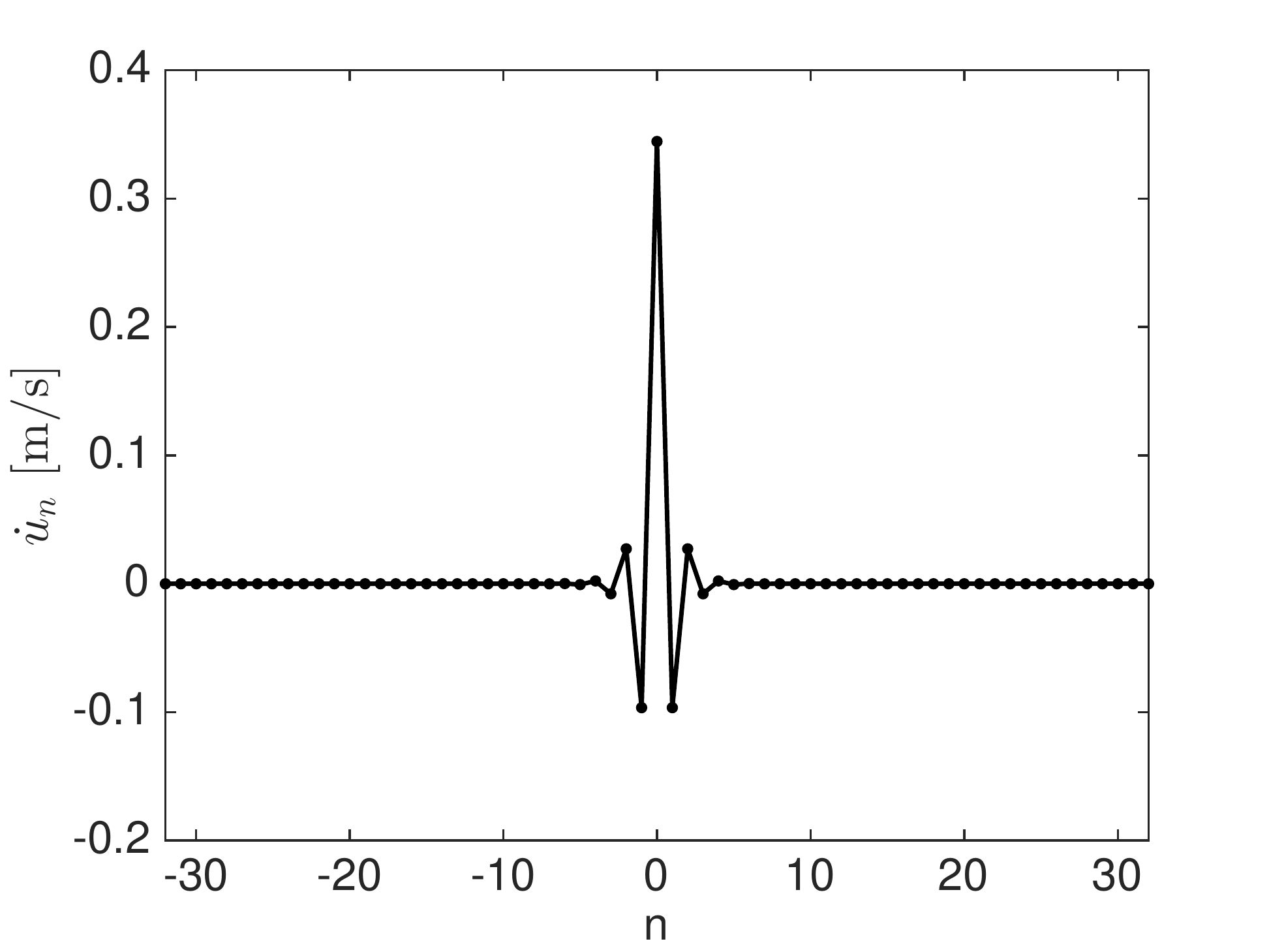} 
  \end{tabular}
\caption{
 Breather solution of Eq.~\eqref{finite} with $\eta=a=0$. We show the displacements of the solution in (a) and their velocities in (b).
}
\label{fig:profiles}
\end{figure}

For experimental  and numerical approaches, we consider a chain of
$N$ (where $N$ is odd) magnets that we arrange as a lattice confined within a distance $L\in\R$ with fixed boundary conditions (i.e., $u_{-\frac{N+1}{2}} = u_{\frac{N+1}{2}} = 0$).  Under these conditions, the equilibrium distance
between magnets $n-1$ and $n$ depends on $n$. 
The $N+1$ equilibrium distances $\delta_{0,n}$ (with $n \in \{-\frac{N-1}{2}, \ldots, \frac{N+1}{2}\}$) satisfy 
$$ \displaystyle L = \sum_{n=-\frac{N-1}{2}}^{\frac{N+1}{2}} \delta_{0,n}$$
and the following $N$ equations:

\begin{equation} \label{cond}
	0 = \sum_{j = -\frac{N+1}{2}}^{n-1}   \left(  \sum_{i = j+1}^{n} \delta_{0,i } \right)^{p} -  \sum_{j = n+1}^{\frac{N+1}{2}}   \left(  \sum_{i = n+1}^{j} \delta_{0,i }  \right)^{p}\,, \qquad n \in \{-\frac{N-1}{2}, \ldots, \frac{N-1}{2}\} \,.
\end{equation}
Thus, for a finite chain, we obtain the following $N$ equations of motion:
\begin{align} \label{finite}
	 M_n \ddot{u}_n &=  \sum_{j = -\frac{N+1}{2}}^{n-1}  A \left(  \sum_{i = j+1}^{n} [ \delta_{0,i } ] + u_n - u_j \right)^{p} -  \sum_{j = n+1}^{\frac{N+1}{2}}  A \left(  \sum_{i = n+1}^{j}[ \delta_{0,i } ]+ u_j - u_n \right)^{p} - \eta \dot{u}_n \,, \qquad n \in
	  \{-\frac{N-1}{2}, \ldots, \frac{N-1}{2}\} \, , \\
 u_{-\frac{N+1}{2}}(t) &= a \, \sin( 2 \pi f_b t) \,, \notag \\ 
 	u_{\frac{N+1}{2}}(t) &= 0\,. \notag
\end{align}

For an infinite lattice (i.e. in the limit $N \rightarrow \infty$) the equilibrium distances are constant with respect to lattice site. This is easily verified by substituting $\delta_{0,n} = \delta_0$ into Eq.~\eqref{cond},
\begin{eqnarray} \label{cond2}
	 \sum_{j=-\infty}^{n-1}   \left(     (n - j) \delta_0   \right)^{p} -  \sum_{j = n+1}^{\infty}   \left(    (j - n)\delta_0  \right)^{p}   &=& \\
	 \sum_{j=1 - n}^{\infty}   \left(     (j + n)    \right)^{p} -  \sum_{j = n+1}^{\infty}   \left(    (j - n)  \right)^{p}   &=& \\
 \sum_{k=1}^{\infty}   \left(    k   \right)^{p} -  \sum_{\ell= 1}^{\infty}   \left(    \ell \right)^{p}  &=& 0 
\end{eqnarray}
where new indices where defined $k = j + n$ and $\ell = j -n$. Substituting $\delta_{0,n} = \delta_0$ into Eq.~\eqref{finite} and redefining indices once again leads to Eq.~(1) in the main text, which is 
valid for an infinite lattice.

We find time-periodic solutions of Eq.~\eqref{finite} with period $T$ by numerically computing roots $x^0$ of the map $f(x^0) = x^0 - \tilde{x}^0(T)$, where $x^0$ is the initial value of Eq.~\eqref{finite} and 
$\tilde{x}^0(T)$
is the solution at time $T$ of Eq.~\eqref{finite} with initial value $x^0$. See \cite{Flach2007} for details. We numerically integrate Eq.~\eqref{finite} with an adaptive-size Runge--Kutta
method. We use the linearization of \eqref{finite} to determine our initial guess for the Newton iterations.  We show a numerical solution with a breather 
frequency $f_b \approx 5.62$ Hz in Fig.~\ref{fig:profiles}.

\end{document}